\documentclass{article}
\usepackage{spconf,amsmath,graphicx}

\usepackage{enumitem}
\usepackage{multirow}
\usepackage{graphicx}
\setlist{nosep, leftmargin=14pt}

\usepackage{mwe} 

\title{SynthMix: Mixing up Aligned Synthesis for Medical Cross-Modality Domain Adaptation}
\name{Xinwen Zhang, Chaoyi Zhang, Dongnan Liu, Qianbi Yu, and Weidong Cai}
\address{School of Computer Science, University of Sydney, Australia}
\begin{document}
%
\maketitle
\begin{abstract}
The adversarial methods showed advanced performance by producing synthetic images to mitigate the domain shift, a common problem due to the hardship of acquiring labelled data in medical field. Most existing studies focus on modifying the network architecture, but little has worked on the GAN training strategy. In this work, we propose SynthMix, an add-on module with a natural yet effective training policy that can promote synthetic quality without altering the network architecture. 
Following the adversarial philosophy of GAN, we designed a mix-up synthesis scheme termed SynthMix. It coherently mixed up aligned images of real and synthetic samples to stimulate the generation of fine-grained features, examined by an associated Inspector for the domain-specific details. 
We evaluated our method on two segmentation benchmarks among three publicly available datasets, where our method showed a significant performance gain compared with existing state-of-the-art approaches. The code will be made publicly available. 
\end{abstract}


\section{Introduction}
The deep learning methods have shown great success in multiple medical image analysis tasks~\cite{unet,deeplearningformedical}. 
Related medical data often shares similarities due to standard human anatomy; thus, information can be transferred across different modalities, such as computed tomography (CT) and magnetic resonance imaging (MRI), to recover exhaustive data acquisition \cite{guan2021domainsurvey}. However, a domain shift between source and target distribution would cause performance degradation when a source-trained model is directly applied on target data~\cite{domainshiftsurvey}. Therefore, unsupervised domain adaptation (UDA) is widely studied to alleviate the domain gap, compensate for the lack of target labels, and best leverage existing resources~\cite{synseg,domainshiftsurvey,liu2022decompose}. 

The adversarial learning regime utilizes the acquired source images to generate target-like images and trains an adversary that discriminates between real and generated samples for UDA. The mini-max game reaches for equilibrium and ultimately achieves an image-level cross-domain transformation. The synthetic images would be used along with source labels to train a target domain algorithm~\cite{cyclegan,synseg,liu2020unsupervised,Yu2022UnsupervisedDA}. 

Mixup as a data augmentation technique has recently proven to boost performance in Deep Learning solutions. It provides a soft transition that diversifies the training samples and helps the model has a more accurate decision boundary~\cite{mixup}. Cut-out~\cite{cutout} randomly replaces a patch of the training image with a black patch, forcing the model to focus on the image's less discriminative part. Cut-mix~\cite{yun2019cutmix} takes advantage of both by replacing a patch of a training image with a cut patch from another training sample. This proposed method fully leverages the training pixels and preserves the regularising merits of regional dropout.

Despite the success of adversarial methods in cross-modality domain adaptation studies, we hypothesize that the generator's potential is suppressed because its opponent, the discriminator, reviews the synthetic product at an image-scale level. Such a design gives the generator little motivation to attend to details and make distinctive local changes—this compromises detailed feature quality, leading to a performance drop in segmentation.

This work presents a novel mix-up scheme for synthetic image alignment to enhance adversarial methods in cross-modality adaptation between CT and MR. Our method promotes the generator's ability to synthesize cross-modality images of better quality, with more fine-grained source domain-specific anatomical traits being retained and more distinguishable target domain-specific characteristics being generated. Our technical contribution includes the following: (a) We propose a spatial mix-up strategy termed SynthMix, that combines two content-aligned yet context-different images to enhance cross-modality synthesis; and (b) We constructed a Mixup Inspector that reasons the global information and determines whether a local patch belongs to the source or target domain to accelerate the training of SynthMix.

\section{Methods}
\begin{figure}[t]
    \centering
    \includegraphics[width=\linewidth]{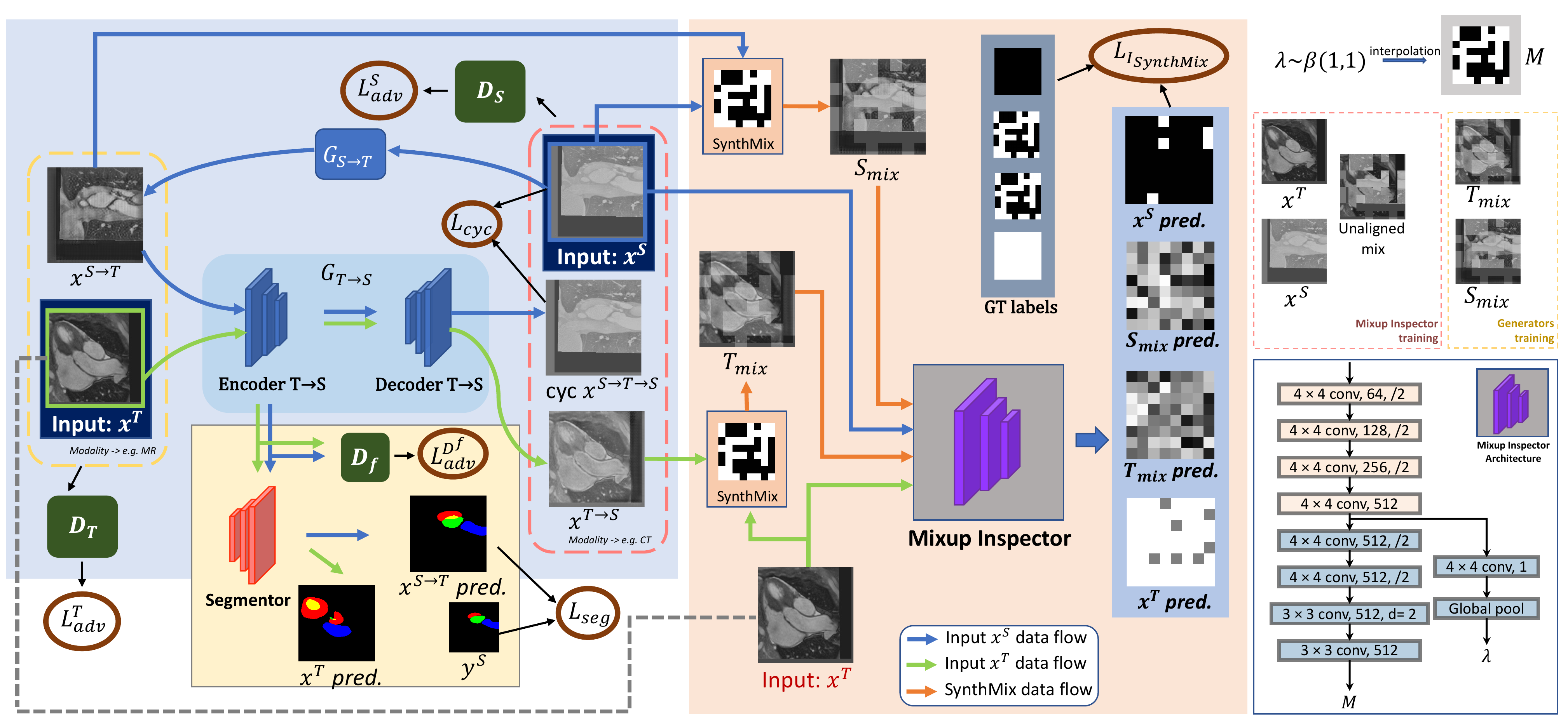}
    \caption{An overview of the proposed method. The inputs S and T (marked in navy blue) go through a generative path for image translation, a SynthMix path for mix-up augmentation, and a segmentation path for semantic label prediction. Zoom in for details.}
    \label{fig:overview}
\end{figure}

Let $x^S$ and $x^T$ denote data in the source domain and target domain, and  $y^S$ denote a segmentation label for source data. The ultimate aim for UDA is to train a segmentor that could predict $y^T$. This objective can be studied through several stages: GAN-based image synthesis to perform the source-to-target translation and to train a segmentor applicable in the target domain (Sec.~\ref{image_syntheis}), training via a SynthMix scheme (Sec.~\ref{synthmix}), and designing an associated Inspector (Sec.~\ref{inspector}). An overview of the model is shown in Fig.~\ref{fig:overview}.

\subsection{Domain Transformation through Image Synthesis}\label{image_syntheis}
Our framework is established on the SIFA architecture~\cite{sifa}. Like CycleGAN~\cite{cyclegan}, SIFA employed two sets of generative adversarial networks to achieve cross-modality image synthesis. Given unpaired source images ($x^S$) and target images ($x^T$), the networks produced source images in target appearances ($x^{S\rightarrow T}$) and target images in source appearances ($x^{T\rightarrow S}$). The generators and discriminators optimized on the adversarial losses $L^S_{adv}$ and $L^T_{adv}$. The cycle consistency loss $L_{cyc}$ was enforced to maintain stability and guide image-scale domain adaptation. An encoder for the target domain was extracted from the target generator $G_{T \rightarrow S}$, followed by a convolutional layer to complete semantic segmentation. An adversarial loss $L^{D_f}_{adv}$ was also included to further align features extracted from the encoder. The overall objective is:
\begin{equation} 
\begin{split}
L = & L^T_{adv}(G_{S \rightarrow T}, D_T) + L^S_{adv}(G_{T \rightarrow S}, D_S) + \\&\lambda_{cyc}L_{cyc}(G_{S \rightarrow T},G_{T \rightarrow S})   + \lambda_{seg}L_{seg}(\mathtt{Seg}) \\ &+ \lambda_{adv}L^{D_f}_{adv}(\mathtt{Seg}, D_f),
\end{split}
\end{equation}

where $G_{S \rightarrow T}$, $G_{T \rightarrow S}$, $D_S$, $D_T$, $\mathtt{Seg}$, $D_f$ are source generator, target generator, source discriminator, target discriminator, target segmentor, feature discriminator, and $\lambda_{cyc}, \lambda_{seg}, \lambda_{adv}$ are trade-off parameters set as {10, 0.1, 0.1} respectively.
The framework employed two discriminators $D_S, D_T$ to supervise image synthesis in their respective domains by distinguishing the domain-specific features. 
For instance, $D_S$ learned to identify source images as authentic and reject $x^{S\rightarrow T}$ images as fakes where they were labelled with scalars 1 and 0 respectively, likewise for $D_T$. 
The discriminators are vital for the network as they are the motivations that compel each generator to perform appearance transformation independently. Hence, they were preserved in our work for the same purpose. Furthermore, we note the limitations of $D_S, D_T$: (1) they only learn the characteristics of one specific domain, (2) they give an entire image a uniform domain label. This results in generators predominantly focusing on the most discriminative features and, consequently, synthesizing delicate features that have less domain distinctive characteristics with poor quality.

\subsection{SynthMix Design}\label{synthmix}
During training, an initial random binary mask of size $k\times k$ was generated, whose occurrences of zeros and ones were controlled by a probability ratio $\lambda$. This mask would next be interpolated to reach the desired mix-up mask $\mathbf{M}$ of matching input size $256\times256$, guiding our SynthMix procedure described below. The SynthMix procedure of the given two samples is defined as:
\begin{gather}
    \tilde{x} = \mathbf{M} \odot x_T+(\mathbf{1}-\mathbf{M})\odot x_S, \\
    \tilde{y} = \mathbf{M} \odot y_T+(\mathbf{1}-\mathbf{M})\odot y_S,
\end{gather}
where $\mathbf{M}$ is the interpolated mask deciding at pixel which image to draw from, $\mathbf{1}$ is a mask of $256\times256$ filled with ones, and $\odot$ is element-wise multiplication. 
Given source $x^S$, target $x^T$, $x^{S\rightarrow T}$ (synthetic target image from a source image), and $x^{T\rightarrow S}$ images, at each training iteration, SynthMix saw 5 kinds of input images: the unmixed source images, the unmixed target images, the unaligned mix of the source and target images, the aligned mixed $S_{mix}$ images (source and $x^{S\rightarrow T}$), and the aligned mixed $T_{mix}$ images (target and $x^{T\rightarrow S}$). The former three were used to train the $I_{SynthMix}$ while the rest were used to optimize the generators $G_{S \rightarrow T}$ and $G_{T \rightarrow S}$. All selected input compositions preserved partial or complete authentic samples from either source or target domain to maintain the integrity of the $I_{SynthMix}$ results.

In SynthMix operation, an image and its cross-domain synthetic fake were mixed following the protocol and processed by the Mixup Inspector to output a map differentiating the domains of each location. The Inspector, trained with real source and target samples and mixed samples of both, learned to classify the domain of each location from a classification loss $L_{cls}$. With the Inspector's knowledge of the features and characteristics of both domains, the generators had to produce better details and locally distinctive features to pass the Inspector. For example, the $G_{S\rightarrow T}$ must generate on par target appearance images $x^{S\rightarrow T}$ which stands out as target patches in the semantically coherent mixed image $S_{mix}$ and vice versa for $G_{T \rightarrow S}$. The generators were optimized on an adversarial loss $L^{mix}_{adv}$.

A significant difference that distinguishes our method and other mix-up strategies is that we mixed up two images with the same semantic contents for generator training. The mixed image containing generator products would be assessed by the $I_{SynthMix}$, which we train to learn both the source and target characteristics. We found that spatially mixing up an image of the same contents would force the generators $G_{S \rightarrow T}, G_{T \rightarrow S}$ to produce refined details in the synthetic images to fool $I_{SynthMix}$, further fulfilling the potential of the generator. Our mix-up approach helped the generator to achieve better domain transformation at no additional data cost, improving the efficiency of data use.

The training objective for SynthMix is:
\begin{equation} 
\begin{split}
L_{I_{SynthMix}} & =  \lambda_{I}L^{mix}_{adv}(G_{S \rightarrow T}, G_{T \rightarrow S}, I)+ \lambda_{I}L_{cls}(I),
\end{split}
\end{equation}
where $\lambda_{I}$ is a hyper-parameter set at 0.1.


\subsection{Mixup Inspector}\label{inspector}
To monitor the training of SynthMix, we propose an associated Mixup Inspector $I_{SynthMix}$ to enhance the supervision. It takes as input an image of size $256\times256$ and produces a $k\times k$ domain mask, whose entries denote the domain classifications of their containing local patches of size $\frac{256}{k}\times\frac{256}{k}$. Extending the PatchGAN~\cite{pix2pix} to an encoder-decoder architecture, this Inspector $I_{SynthMix}$ is designed to learn domain-specific characteristics on images from both domains. 
For an extra supervision that helps the model converge, an additional branch is added to pass these feature maps through a 4×4 convolution and a global average pooling layers, which is later supervised by the domain-specific scores averaged among $\mathbf{M}$. 
SynthMix processed samples with randomly mixed labels, forcing the Inspector to analyze the image at the locality and ultimately learn the domains at the pre-designated disjoint patches.

\section{Experiments and Results}
We assessed our SynthMix method on two segmentation benchmarks, followed by the comparisons of segmentation results with SOTA approaches. The experiment settings and results are reported in Section~\ref{sec: dataset} and Section~\ref{sec: results} respectively, followed by the ablation studies of our method presented in Section~\ref{sec:ablation}. 

\subsection{Experiment Settings} \label{sec: dataset}

Following SIFA~\cite{sifa}, we evaluated our unpaired CT-MR adaptation model on two segmentation benchmarks, including the segmentation of cardiac structures and abdominal organs, among three publicly available biomedical segmentation datasets. More specifically, we adopted Multi-Modality Whole Heart Segmentation (\textbf{MMWHS})~\cite{MMWHS} dataset for cardiac structure segmentation. Following [1], we merged the Combined Healthy Abdominal Organ Segmentation (CHAOS) dataset~\cite{CHAOS2021} and the Multi-Atlas Labeling Beyond the Cranial Vault challenge (Multi-Atlas Labeling) dataset~\cite{beyondcranial}, into an Abdominal Organ Segmentation (\textbf{AOS}) benchmark dataset for evaluation.

The model was evaluated with two standard metrics: the Dice coefficient and the average symmetric surface distance (ASSD). The Dice coefficient calculates the similarity of prediction and reference masks by promoting overlapping volumes and punishing non-overlapping volumes, while the ASSD measures the distances between the surfaces of prediction and the ground truth masks. 

\subsection{Medical Image Segmentation} \label{sec: results}

\begin{table}[!b]
\caption{Comparison of domain adaptation (Cardiac MR-to-CT) performance on MMWHS dataset.}\label{tab1:MMWHS}
\setlength{\tabcolsep}{0.6em}
\begin{tabular*}{\linewidth}{c|ccccc}

\multirow{2}{*}{Method} & \multicolumn{1}{l}{\scriptsize AA} & {\scriptsize LAC}  & {\scriptsize LVC}  & {\scriptsize MYO}  & {\scriptsize Avg}  
                          \\ \cline{2-6} 
                  & \multicolumn{5}{c}{Dice (\%) $\uparrow$}\\  \hline
CycleGAN~\cite{cyclegan}   & 73.8                   & 75.7 & 52.3 & 28.7 & 57.6 \\
SIFA~\cite{sifa}        & 81.3                   & 79.5 & 73.8 & 61.6 & 74.1 \\
SIFA+Mixup~\cite{mixup}        & 86.1                   & 82.5 & 79.8 & 71.5 & 80.0 \\
SIFA+CutMix~\cite{yun2019cutmix}       & 83.3                   & 85.4 & \textbf{86.4} & 67.6 & 80.7 \\
\textbf{Ours}             &     \textbf{87.2}             &  \textbf{88.5}    &    82.4 &   \textbf{71.8}   &  \textbf{82.5}        \\ \hline 


& \multicolumn{5}{c}{ASSD (\%) $\downarrow$} \\ \cline{2-6}
CycleGAN~\cite{cyclegan}   &  11.5 & 13.6  & 9.2 & 8.8 & 10.8 \\
SIFA~\cite{sifa}        & 7.9 & 6.2  & 5.5 & 8.5 & 7.0 \\
SIFA+Mixup~\cite{mixup}        &  5.5 & 3.8  & 3.8 & 4.6 & 4.4 \\
SIFA+CutMix~\cite{yun2019cutmix}       &  9.2 & 4.0  & \textbf{3.0} & 4.2 & 5.1 \\
\textbf{Ours}             &      \textbf{5.4}   &  \textbf{3.2}    &  3.3  &  \textbf{3.5}   & \textbf{3.8}     \\ \hline 
\end{tabular*}
\end{table}

We compared our method with several existing SOTA approaches on UDA, including SIFA~\cite{sifa}. Table~\ref{tab1:MMWHS}, Table~\ref{tab2:MMWHS}, and Table~\ref{tab2:AOS} provide a summary of performance differences on MMWHS and AOS datasets respectively.
We performed MR-to-CT transformation for the unlabeled cardiac CT segmentation and in the CT-to-MR direction for MR segmentation to demonstrate the pipeline's ability to perform domain adaptation in both directions. We also evaluated the method on the AOS dataset in CT-to-MR direction for abdominal organ segmentation to assess the method's robustness to subject changes.

\begin{table}[b!]
\caption{Comparison of domain adaptation (Cardiac CT-to-MR) performance on MMWHS dataset.}\label{tab2:MMWHS}
\setlength{\tabcolsep}{0.55em}
\begin{tabular*}{\linewidth}{c|ccccc}

\multirow{2}{*}{Method} &\multicolumn{1}{l}{\scriptsize AA} & {\scriptsize LAC}  & {\scriptsize LVC}  & {\scriptsize MYO}  & {\scriptsize Avg}  
\\ \cline{2-6} 
                  & 
                  \multicolumn{5}{c}{Dice (\%) $\uparrow$}\\ \hline
CycleGAN~\cite{cyclegan}   & 64.3                   & 30.7 & 65.0 & 43.0 & 50.7  \\
SIFA~\cite{sifa}        & 65.3                   & 62.3 & 78.9 & 47.3 & 63.4 \\
SIFA+Mixup~\cite{sifa,mixup}        & 67.0                  & 65.9 & 77.6 & 41.2 & 62.9 \\
SIFA+CutMix~\cite{sifa,yun2019cutmix}        & 63.0       & 66.0 & 81.0 & 47.6 & 64.4 \\
\textbf{Ours}             &     \textbf{70.0}             &  \textbf{68.7}    &    \textbf{83.0}  &   \textbf{48.3}   &  \textbf{67.5}       \\ \hline
   & \multicolumn{5}{c}{ASSD $\downarrow$}  
                 \\  \cline{2-6}
CycleGAN~\cite{cyclegan}    & 5.8 & 9.8  & 6.0 & 5.0 & 6.6 \\
SIFA~\cite{sifa}         & 7.3 & 7.4  & 3.8 & 4.4 & 5.7 \\
SIFA+Mixup~\cite{sifa,mixup}         & \textbf{5.2} & 6.9  & 3.9 & 4.5 & 5.1 \\
SIFA+CutMix~\cite{sifa,yun2019cutmix}        & 6.3 & \textbf{4.9}  & \textbf{3.2} & \textbf{4.0} & 4.6 \\
\textbf{Ours}              &  5.9   &  5.1    &  3.3   &  4.1  & \textbf{4.6}\\ \hline

\end{tabular*}
\end{table}

\begin{figure}[t]
    \centering
    \includegraphics[width=\linewidth]{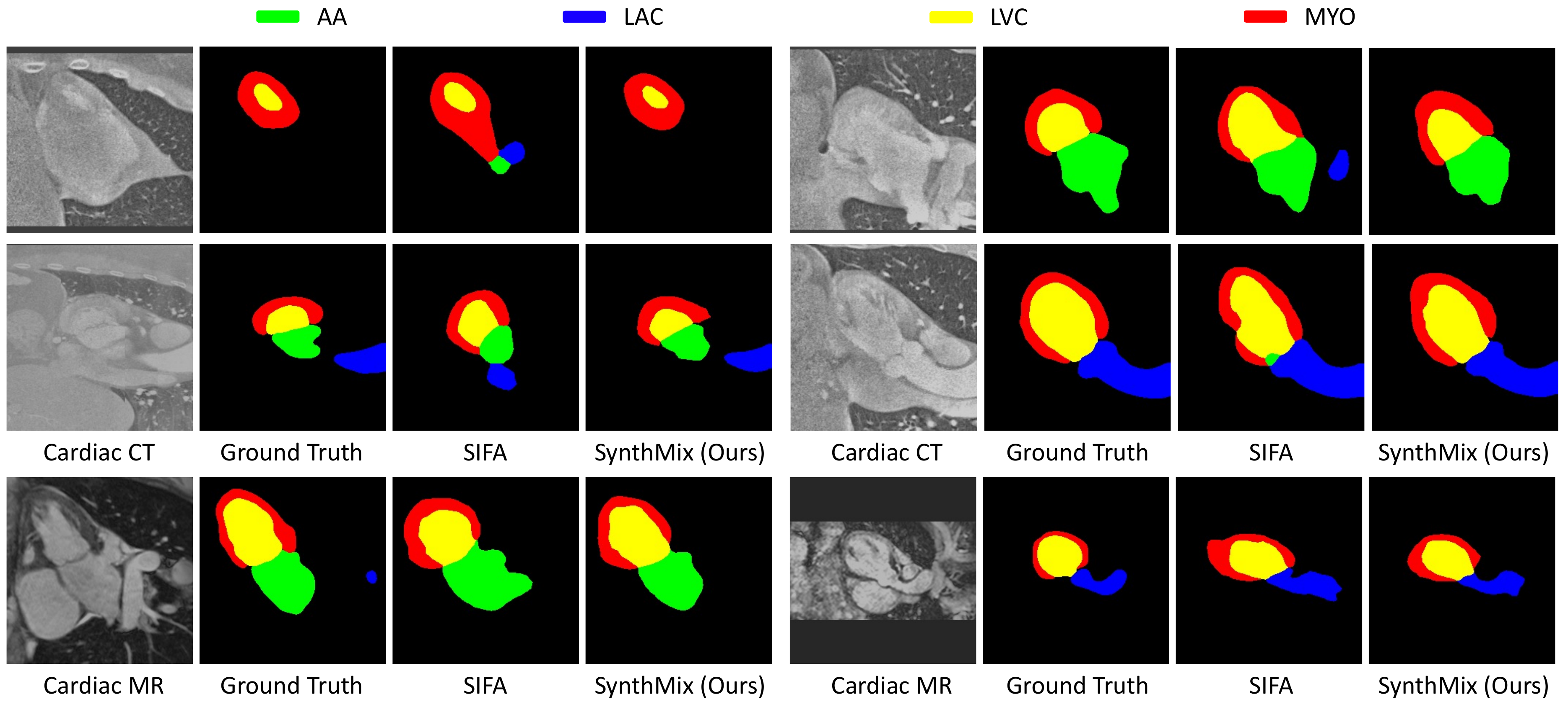}
    \caption{Qualitative comparison results for medical image segmentation. Top two rows denote adapted CT results and bottom row presents adapted MR results. Zoom in for details.}
    \label{fig:seg}
\end{figure}

\begin{table}[b!]
\caption{Comparison of CT-to-MR domain adaptation performance on AOS dataset.}\label{tab2:AOS}
\setlength{\tabcolsep}{0.6em}
\begin{tabular}{c|ccccc}

\multirow{2}{*}{Method} &
\multicolumn{1}{l}{\scriptsize Liver} & {\tiny R Kidney}  & {\tiny L Kidney}  & {\scriptsize Spleen}  & {\scriptsize Avg}
\\ \cline{2-6} 
                & \multicolumn{5}{c}{Dice (\%) $\uparrow$} 
                  \\ \hline
CycleGAN~\cite{cyclegan}   & \textbf{88.8}                 & 87.3 & 76.8 & 79.4 & 83.1 \\
SIFA~\cite{sifa}        & 90.0                   & \textbf{89.1} & 80.2 & 82.3 & \textbf{85.4} \\
SIFA+Mixup~\cite{mixup}        & 86.2             & 84.5 & 82.1 & 76.0 & 82.2  \\
SIFA+CutMix~\cite{yun2019cutmix}        & 83.7                   & 86.7 & 77.6 & 76.0 & 81.0 \\
\textbf{Ours}              &       86.6                 & 88.8     &  \textbf{82.7}    &   \textbf{82.9}  &   85.2   \\ \hline
    & \multicolumn{5}{c}{ASSD $\downarrow$}     \\ \cline{2-6}
CycleGAN~\cite{cyclegan}   & 2.0 & 3.2  & 1.9 & 2.6 & 2.4 \\
SIFA~\cite{sifa} & 1.5 & 0.6  & 1.5 & 2.4 & 1.5 \\
SIFA+Mixup~\cite{mixup} & 0.7 & 1.7  & 0.5 & 1.2 & 1.0 \\
SIFA+CutMix~\cite{yun2019cutmix}  & 0.5 & 0.5  & 0.6 & 1.2 & 0.7 \\
\textbf{Ours}        &  \textbf{0.4 }    & \textbf{0.2} &  \textbf{0.5} & \textbf{1.1} &  \textbf{0.6}   \\
\hline

\end{tabular}
\end{table}

The table shows the performance of existing SOTA adversarial UDA methods: CycleGAN~\cite{cycada} is one of the foundation studies achieving adversarial domain adaptation via image translation,
while SIFA~\cite{sifa} synergizes $G_{T \rightarrow S}$ and $\mathtt{Seg}$ by sharing encoders, along with an alternative feature alignment. To compare SynthMix with existing Mixup methods, we used Mixup~\cite{mixup} and CutMix~\cite{yun2019cutmix} as the mix-up protocols and their default ResNet as the auxiliary domain inspector to perform similar data augmentation for the SIFA framework. 
All these methods were trained on images from both domains (with source labels only), and their target segmentors were evaluated by testing images from the target domain.

It is evident that our SynthMix outperforms others on the cardiac MMWHS dataset, where it ranked \textit{1st} in most categories with clear margins in both cross-modality transforming directions. The averages of Dice and ASSD have an 8.4\%$\uparrow$ and 3.2$\downarrow$ (MR-to-CT) and a 4.1\%$\uparrow$ and 1.1$\downarrow$ (CT-to-MR) performance gain over SIFA, which could be verified by the qualitative comparisons with SOTA on MMWHS as shown in Fig.~\ref{fig:seg}. To demonstrate the robustness of our method, we also validated our network on the AOS dataset in the CT-to-MR direction. Table~\ref{tab2:AOS} shows that SynthMix achieves significant and consistent improvements in ASSD over all classes, while comparable results were also maintained on Dice. We also conducted a 2-tailed t-test on SynthMix with current SOTA SIFA and achieved a p-value smaller than 0.01, showing significant improvement. Furthermore, the comparison drawn between our method and other mix-up augmentations showed the immediate enhancement of our mix-up framework for generative UDA and proved the consistent improvement of SynthMix over other mix-up protocols in all tasks without additional tuning.

\subsection{Ablation Studies}\label{sec:ablation}
We conducted an ablation study of essential components in the SynthMix framework, as shown in Table~\ref{tab:abs}. Following the official TensorFlow SIFA repository \footnote[1]{https://github.com/cchen-cc/SIFA}, we implemented a baseline in Pytorch (row 1). The second row shows an ablative setting when both adversarial discriminators $D_S$ and $D_T$ in SIFA are replaced with a single Mixup Inspector. Based on the inferior results, we notice that the original discriminators should be kept and working together with the Inspector for SynthMix. 
Next, rows 4 and 5 present two alternative SynthMix designs on mix-up resolution. We modified the initial binary mask size $k$ and the decoder layers of the Mixup Inspector to produce matching $k\times k$ output. 
In the final row, we present SynthMix. The input for the Inspector is selected as the real source, target, and unaligned mix, while the SynthMix resolution is set at 8, which gives the best performance.

\begin{table}[t]
    \centering
    \caption{Ablation studies on MR-to-CT setting on MMWHS dataset.}\label{tab:abs}
    \begin{tabular}{l|cc}
    \hline
          \multicolumn{1}{c|}{{Ablative Setting}} & Dice & ASSD \\
         \hline
         Baseline $I_{SynthMix}$ & 78.1 & 6.0 \\
         
         Model 0: No $D_{S}$, $D_{T}$ & 77.4 & 5.7  \\
         \hline
         Model 1: SynthMix resolution $k$= 4 & 79.9 & 5.6 \\
         Model 2: SynthMix resolution $k$= 32 & 80.9  & 4.5  \\
         \hline
        \textbf{Ours: }$\mathbf{k = 8}$ & 82.5 & 3.8
         \\
         \hline
    \end{tabular}
    \label{tab:my_label}
\end{table}

\section{Conclusion}

In this work, we proposed a novel framework termed SynthMix for UDA segmentation based on aligning real and synthetic image Mixup for medical data. We designed a spatial mix-up data augmentation technique and investigated its operations, followed by constructing a novel network to supervise our mix-up scheme. The experimental results on various cross-modality medical segmentation tasks demonstrate the superiority of SynthMix over existing SOTA approaches. 

\bibliographystyle{IEEEbib}
\bibliography{strings,refs}

\end{document}